# Formalism for Anatomy-Independent Projection and Optimization of Transcranial Magnetic Stimulation Coils

Max Koehler and Stefan M. Goetz

*Abstract*—Transcranial magnetic stimulation (TMS) is a popular method for the noninvasive stimulation of neurons in the brain. It has become a standard instrument in experimental brain research and is approved for a range of diagnostic and therapeutic applications. These applications require appropriately shaped coils. Various applications have been established or approved for specific coil designs with their corresponding spatial electric field distributions. However, the specific coil implementation may no longer be appropriate from the perspective of material and manufacturing opportunities or considering the latest understanding of how to achieve induced electric fields in the head most efficiently. Furthermore, in some cases, field measurements of coils with unknown winding or a user-defined field are available and require an actual implementation. Similar applications exist for magnetic resonance imaging coils.

This work aims at introducing a formalism that is completely free from heuristics, iterative optimization, and ad-hoc or manual steps to form practical stimulation coils with a winding consisting of individual turns to either equivalently match an existing coil or produce a given field. The target coil can reside on practically any sufficiently large or closed surface adjacent to or around the head. The method derives an equivalent field through vector projection. In contrast to other coil design or optimization approaches recently presented, the procedure is an explicit forward Hilbert-space vector projection or basis change. For demonstration, we map a commercial figure-of-eight coil as one of the most widely used devices and a more intricate coil recently approved clinically for addiction treatment (H4) onto a bent surface close to the head for highest efficiency and lowest field energy. The resulting projections are within $\leq 4\%$ of the target field and reduce the necessary pulse energy by more than 40%.

*Keywords:* Coil design, coil equivalencies, coil simulation, magnetic vector potential, modal decomposition, vector projection, transcranial magnetic stimulation (TMS).

## I. Introduction

MAGNETIC stimulation uses strong brief magnetic pulses to induce currents into tissue and activate nerves or muscles [1-5]. Its transcranial form called TMS allows the writing of artificial signals across the skull into neurons and neural circuits inside the brain [6,7]. How neuronal circuits process endogenous signals can be modulated by the stimulation/modulation with certain pulse rhythms and patterns [8]. TMS has become an essential tool in experimental brain research and is widely used in medical diagnosis and treatment [9-12]. It is, for example, cleared in various countries for the treatment of depression, bipolar disorder, obsessive-compulsive disorder, smoking addiction, and migraine as well as for various diagnostic procedures and cortical mapping [13-26]. TMS is also under investigation for many other disorders [6, 9, 27-29].

The spatial distribution of the induced electric field determines which brain circuits are activated. This distribution already strongly depends on the coil design, i.e., the shape of the conductors in the stimulation coil, and the exact position of the coil during stimulation [30]. Device manufacturers and researchers have developed a variety of different coils for TMS [31].

Focal coils, such as figure-of-eight coils, allow the activation of smaller targets and circuits down to individual muscle representations [32]. Replacing the coil in a procedure can influence the outcome and its efficacy, even if the coils appear very similar [33]. Accordingly, various clinical applications are furthermore approved for very specific coils [14]. Replacing the coil with another can void the approval.

Such procedures would benefit from a coil with the same field as a specific device but generated in a better, e.g., more efficient way instead of novel coils with novel features with unknown physiology and void clinical approval. This aspect might be a key reason, why the majority of suggested coil designs in the literature have never been translated into actual use [31].

In other cases, the specific approved coil for a procedure is not appropriate or practical. It may often have the wrong size. Bent shapes, for instance, have to go around the head but might be too small to fit or too large, losing efficiency [73]. Some coils might mechanically interfere with other equipment, such as electroencephalography electrodes, near-infrared transducers, or implants, and would require a modified coil with equal field distribution that accounts for the specific constraints. Furthermore, the specific approved coil may for historic reasons just have elements that are far from the head as the original design stems from a time when efficiency and coil heating did not matter yet due to primarily single-pulse operation, and/or the understanding of design rules for efficient coils was not known yet. Furthermore, complicated and expensive manual manufacturing of an unnecessarily complicated winding was acceptable when only few prototypes were needed. However, the routine clinical use following a certain approved procedure would require the use of the very same coil design with its approved field shape, no matter how impractical or substandard it might be in the light of latest technology or manufacturing.

As such, many early coils contain elements that protrude from the head or have some distance from the head [32]. Even some newer designs use elements that do not touch the head to reduce and spread out the induced electric field, e.g., outside the target area to achieve higher focality [35]. However, it is known that protruding elements reduce the electromagnetic coupling to the head while increasing heating [36]. On the other hand, it is known by now that any electromagnetic field inside a confined volume, such as a sphere or a head, can in principle also be generated by currents on any single-layer surface closed around the target volume. Open surfaces and even planes can approximate a closed surface if they are sufficiently large [37-40].

In some other cases, the desired field of a coil is known, e.g., measured, but not yet an actual coil implementation. While many coils use a potted inaccessible winding, there is substantial effort in the field to catalog and measure field profiles for documentation and models without knowledge of the internal windings [41-43].

Previous attempts trying to optimize coils or to match a given electromagnetic field contain ad-hoc, heuristic steps or are fully manual [69-71, 73]. Particularly the generation of discrete closed wire paths is a widely manual procedure in all previous reports, which was explicitly criticized before, though still not solved [44, 45, 46, 66]. Instead of such heuristics or manual steps, a transformation or projection guaranteeing mathematical and physical equivalence would be preferable. Previous research has optimized coils, which could in principle also serve for deriving equivalent coils [34, 45, 46, 73]. However, the use of a global optimization framework, typically including search heuristics, does not necessarily guarantee a good match or equivalence and further appears computationally excessive if instead a forward projection transformation were possible.

The problem of coil optimization is also pressing in other fields. In magnetic resonance imaging (MRI), particularly the gradient coils use complicated spatial patterns to achieve accurate field conditions, such as high gradient linearity or low stimulation potential, with low energy consumption [47-51]. The formalistic and technical challenges used for MRI match those in magnetic stimulation. So do plausibly the techniques used for coil optimization [74-77]. However, the methods also suffer from the same limitations.

This article derives a formalism to create coils, or to be more exact, conductor paths on a simple geometric shell to match a magnetic vector potential within a closed volume through vector projection of basis vectors.

## II. Anatomy-Independence For The magnetic Vector Potential

This article aims at finding simple conductor paths fully residing on an arbitrarily shaped shell around or a surface near the head through projection operations only. Since the coil should generate a practically equivalent induced electric field profile to the targeted magnetic field in any patient, i.e., any conceivable head and brain anatomy, we suggest using a precursor of the induced electric field for matching. Previous approaches of optimization focused on the resulting induced electric field, which would require an entire ensemble of realistic head anatomies covering the full bandwidth of possible anatomic variability including abnormalities. The use of an electromagnetic, anatomy-independent precursor as suggested here, however, intrinsically enforces equivalent outcomes compared to the original coil or field for any head anatomy due to the Maxwell equations. The use of a standard anatomy, such as the MNI average, does not appear attractive: there would be no information if the derived coil would maintain equivalence in any other or actual anatomy; furthermore, standard anatomies, such as the MNI head average out and miss many actual features of the brain that shape the field in TMS, such as the strong gyrification. The chosen matched physical quantity is the magnetic vector potential $\mathbf{A}$ inside the spherical region of interest, e.g., a sphere around the brain, which fully determines the magnetic flux density $\mathbf{B}$ everywhere inside through

$$\mathbf{B}(\mathbf{r}) = \mathbf{curl}\, \mathbf{A}(\mathbf{r}) = \nabla \times \mathbf{A}(\mathbf{r}) \tag{1}$$

at the location $\mathbf{r}$ with the curl operator $\mathbf{curl}$ or, in nabla representation, $\nabla \times$ [38]. Although $\mathbf{A}$ and $\mathbf{B}$ are only forced into equivalency inside the shell, the definition in Equation (1) pertains outside and ensures that magnetic flux lines are closed. Thus, the magnetic flux lines forced to equivalency inside the region of interest are closed somewhere outside per

$$\mathbf{div}\, \mathbf{B}(\mathbf{r}) = \nabla \cdot \mathbf{B}(\mathbf{r}) = \nabla \cdot \big(\nabla \times \mathbf{A}(\mathbf{r})\big) = 0, \tag{2}$$

with the divergence operator div or $\nabla$. This relationship is commonly known as Gaussian law in electrodynamics. Outside the region of interest, the original coil's field and the derived, simplified one may substantially deviate without any relevant impact. With sufficiently equal magnetic vector potential and magnetic flux inside the region of interest, the induced electric field inside will again be the same between original and equivalent, no matter which anatomy or object is inside the volume.

The induced electric field can be derived through Faraday's law of induction, which reads

$$\mathbf{E}(\mathbf{r}) = -\frac{\partial}{\partial t}\mathbf{A}(\mathbf{r}) - \mathbf{grad}\,\phi(\mathbf{r}) \tag{3}$$

for the vector potential and the electrical potential distribution $\phi(\mathbf{r})$ in the vector analysis formalism [36, 52]. The electrical potential is generated by the currents flowing in response to the induced electric fields, which in turn accumulate at conductivity changes, e.g., at tissue interfaces or durae, and generate a counter field $-\mathbf{grad}\,\phi(\mathbf{r})$. The first summand of Equation (3) is sometimes called the primary induced electric field

$$\mathbf{E}_{\text{prim}}(\mathbf{r}) = -\frac{\partial}{\partial t}\mathbf{A}(\mathbf{r}), \tag{4}$$

which per definition of the problem statement above has to be the same for the original coil as well as the derived one. It causes an electrical potential $\phi$ at tissue transitions according to

$$\mathbf{div}\big(\sigma(\mathbf{r})\,\mathbf{grad}\,\phi(\mathbf{r})\big) = -\frac{\partial}{\partial t}\mathbf{A}(\mathbf{r}) \cdot \mathbf{grad}\,\sigma(\mathbf{r}) \tag{5}$$

with the conductivity distribution $\sigma(\mathbf{r})$ in space, which describes the anatomy. As all terms will be equal for the original coil and the derived one, including the anatomy representation $\sigma(\mathbf{r})$ for any patient, also the overall induced electric field will be the same for any anatomy through such proper problem formulation. Using the electric field instead of the magnetic vector potential as regularly suggested and done in all previous approaches would only provide anatomy-dependent solutions, represented by the conductivity distribution, resulting in a loss of generality. The magnetic vector potential serves in this work to completely omit the influence of anatomic dependencies.

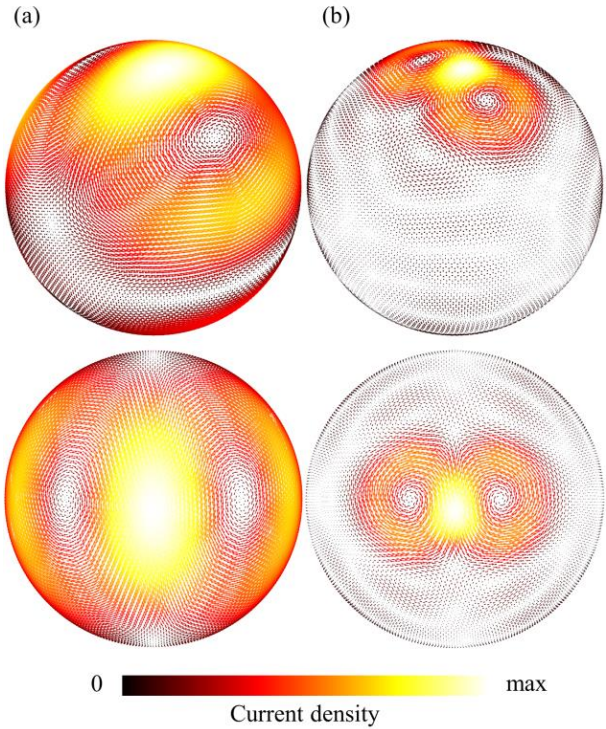

Fig. 1. Perspective (upper row) and top view (lower row) of the continuous current distribution after mapping the D70 Alpha coil to a sphere. (a) Equal current distribution for the first harmonic base vectors up to order 3 and (b) up to order 9. The amplitude is color-coded.

## III. Solution Space

We define the solution space of all possible coils as the current distribution on a shell encompassing the head, specifically a sphere. To describe practically any possible coil residing on the surface, we use a continuous vector-valued current distribution $\mathbf{j}(\mathbf{r})$ at position $\mathbf{r}$, which as the most fundamental constraint only fulfills current continuity per

$$\mathbf{div}\big(\mathbf{j}(\mathbf{r})\big) = 0. \tag{6}$$

An ideal coil is therefore described by the current distribution $\mathbf{j}(\mathbf{r})$ on the surface. Its current distribution can also be represented by a sum of several currents. If the solution space is designed as a complete infinite vector space, e.g., a Banach space, it can be given a vector frame or basis, and every possible solution can be represented as a sum of basis or frame vectors $\mathbf{b}_i(\mathbf{r})$ each scaled appropriately by $\gamma_i$ following

$$\mathbf{j}(\mathbf{r}) = \sum_i \gamma_i \, \mathbf{b}_i(\mathbf{r}). \tag{7}$$

The problem of finding an appropriate coil, therefore, becomes the search for weights or coordinates $\gamma_i$ for the frame or basis vectors. This design still vaguely follows concepts from previous work on coil optimization [45]. Any basis of the current space can be used for further steps.

We deviate from any previous coil optimization approaches from here on by using the magnetic vector potential space and defining a projection operation. Since the electromagnetic induction of such air coils is linear, every sum of currents also entails a sum of corresponding magnetic vector potentials and thus corresponding induced electric fields. Likewise, each frame or basis vector $\mathbf{b}_i(\mathbf{r})$ has its own representation in the Banach space of magnetic vector potentials, forming an isomorphic frame or basis $\{\mathbf{a}_i\}$ there according to

$$\mathbf{a}_i = \mathfrak{U}\{\boldsymbol{b}_i\}, \tag{8}$$

where the functional $\mathfrak{U}$ calculates the magnetic vector potential $\mathbf{a}$ of the spatial current distribution in its argument, e.g., the frame or basis vectors of the coil current space. Due to the linearity of $\mathfrak{U}$, $\mathfrak{U}(\sum_i \gamma_i \, \mathbf{b}_i) = \sum_i \gamma_i \, \mathfrak{U}\{\boldsymbol{b}_i\}$. The magnetic vector potential is derived via Biot–Savart here [52, 53].

We define a bilinear inner product through

$$\langle \mathbf{A}_i, \mathbf{A}_j \rangle \coloneqq \int \mathbf{A}_i(\mathbf{r}) \, \mathbf{A}_j(\mathbf{r}) \, \mathrm{d}^3 r \tag{9}$$

for any vectors $\mathbf{A}_i$, which allows projections of any coil described through its current distribution onto bases of the current space. The inner product turns the Banach-type current space representing any possible coil on the surface into a Hilbert space.

## IV. Frame or Basis

Depending on the chosen solution space, for example, a plane, a spherical shell, or an ellipsoid, the frame or basis must be chosen adequately. Any complete frame or basis of closed loops in the surface for a divergence-free current could serve to span the coil current space. A formulation for general frames in contrast to the prior art allows a relatively simple design of a generator ignoring any orthogonality or even similarity, which may numerically induce coupling, and may enable more appropriate ones, e.g., for better numerical stability, in the future. A rather simple basis for a planar solution space could consist of standing sinusoidal waves of increasing frequency to form the modes of the surface in each direction. For spherical surfaces, their harmonics, for instance, form an orthonormal and complete modal decomposition of closed loops and can therefore represent in linear combinations the current of any coil shape [45]. It would also be conceivable to construct a wide range of alternative bases [54].

Subsequently, we translate the coil frame, e.g., a modal coil basis as suggested above, into the isomorphic vector potential space, from which the primary electric field of every frame or basis vector follows readily.

The basis can be implemented through analytical equations as we will do in our examples or numerically. As known from other fields of numerics, analytical representations, where they exist, can substantially improve stability, particularly if derivatives have to be evaluated. However, as previous work has shown, numerical representations, e.g., as discrete vectors in tessellated surfaces, as an alternative can allow more flexibility for less ordinary basis definitions and more irregular surfaces.

## V. Projection Onto The Frame Or Basis Vectors

Given a general frame $\{\mathbf{b}_i\}$ in the coil current space on the surface with corresponding isomorphic frame $\{\mathbf{a}_i\}$ in the magnetic vector space, both of which are not necessarily linearly independent, the projection of the given magnetic vector potential $\mathbf{A}_{\text{given}}(\mathbf{r})$ inside the region of interest $V_{\text{ROI}}$ of a specific coil to be matched represents the search for a linear combination of magnetic vector potentials of the frame vectors

$$\mathbf{A}_{\text{given}}(\mathbf{r}) - \sum_i \gamma_i \mathbf{a}_i(\mathbf{r}) = \mathbf{A}_{\text{given}}(\mathbf{r}) - \sum_i \gamma_i \mathfrak{U}\{\mathbf{b}_i\}. \quad (10)$$

If Eq. (10) is multiplied with any frame vector $\mathbf{a}_j$ using the above-defined inner product, it generates $j$ equations to be fulfilled per

$$\langle \mathbf{A}_{\text{given}}(\mathbf{r}), \mathbf{a}_j(\mathbf{r}) \rangle - \langle \sum_i \gamma_i \mathbf{a}_i(\mathbf{r}), \mathbf{a}_j(\mathbf{r}) \rangle \; \forall j, \text{ and through}$$

bilinearity

$$\langle \mathbf{A}_{\text{given}}(\mathbf{r}), \mathbf{a}_j(\mathbf{r}) \rangle - \sum_i \gamma_i \underbrace{\langle \mathbf{a}_i(\mathbf{r}), \mathbf{a}_j(\mathbf{r}) \rangle}_{\rho_{ij} :=} \; \forall j. \quad (11)$$

For general not fully linearly independent $\{\mathbf{a}_i\}$, Eq. (11) accordingly forms a matrix–vector equation

$$\sum_i \gamma_i \rho_{ij} = \underbrace{\langle \mathbf{A}_{\text{given}}(\mathbf{r}), \mathbf{a}_j(\mathbf{r}) \rangle}_{\lambda_j :=}, \quad (12)$$

with $\mathbf{P} = (\rho_{ij})$, $\boldsymbol{\gamma} = (\gamma_i)$, and $\boldsymbol{\lambda} = (\lambda_j)$.

For $\{\mathbf{a}_i\}$ only containing those vectors that cannot be fully represented by other $\mathbf{a}_i$, i.e., no redundant frame vectors, turning $\{\mathbf{a}_i\}$ mathematically into an exact frame, the equation has a unique solution for the required vector $\boldsymbol{\gamma}$ as the right-hand side $\boldsymbol{\lambda}$ can be readily evaluated.

As the coil current space is designed isomorphic to the space of corresponding magnetic vector potentials, the equivalent current density $\mathbf{j}_{\text{equiv}}$ follows

$$\mathbf{j}_{\text{equiv}}(\mathbf{r}) = \sum_i \gamma_i \mathbf{b}_i(\mathbf{r}). \quad (13)$$

In the special case of an orthonormal basis, i.e., $\langle \mathbf{a}_i(\mathbf{r}), \mathbf{a}_j(\mathbf{r}) \rangle = \langle \mathbf{b}_i(\mathbf{r}), \mathbf{b}_j(\mathbf{r}) \rangle = \rho_{ij} = \delta_{ij}$ with the Kronecker delta $\delta_{ij}$, the matrix–vector equation in Eq. (12) decouples. Accordingly, the required factors $\gamma_i$ equal the simple projection of the given magnetic vector potential $\mathbf{A}_{\text{given}}(\mathbf{r})$ onto the magnetic vector potentials of each coil current basis vector $\mathbf{a}_i(\mathbf{r}) = \mathfrak{U}\{\mathbf{b}_i\}$ per

$$\gamma_i = \langle \mathbf{A}_{\text{given}}(\mathbf{r}), \mathbf{a}_i(\mathbf{r}) \rangle = \int_{V_{\text{ROI}}} \mathbf{A}_{\text{given}}(\mathbf{r}) \, \mathfrak{U}\{\mathbf{b}_i\} \, \mathrm{d}^3 r. \quad (14)$$

## VI. DISCRETIZATION OF CURRENT DISTRIBUTION INTO A COIL CONDUCTOR PATH AND INDIVIDUAL TURNS

The equivalent current distribution $\mathbf{j}_{\text{equiv}}$ on the surface is continuous and furthermore per the design of the underlying vector space divergence-free; therefore, numerous unconnected closed loops would form. For the implementation into a continuous-wire coil, we discretized the distribution into a single conductor path. The divergence freedom, which current should, in general, fulfill globally, is locally a mathematical and physical challenge as the resulting coil representation should ideally be a single wire path with a start and end, e.g., a spiral, and ideally adjustable inductance. Previous work that discretizes continuous currents for a coil typically generates multiple unconnected closed loops along contour lines first and links them more or less abruptly. This step has in the strict sense ad hoc and usually involves manual or otherwise unjustified steps [44]. The winding generation in previous work, for example, forms several closed loops. The closed loops are then typically opened in one or several locations and interconnected in rather abrupt steps, which form sharp bends which are hard to wind. The positions of these interconnects are left to the discretion of the designer, instead of any consistent physical or mathematical rule. Smoothening the loops into each other through spatial low-pass filters departs from the original current path without consistent control.

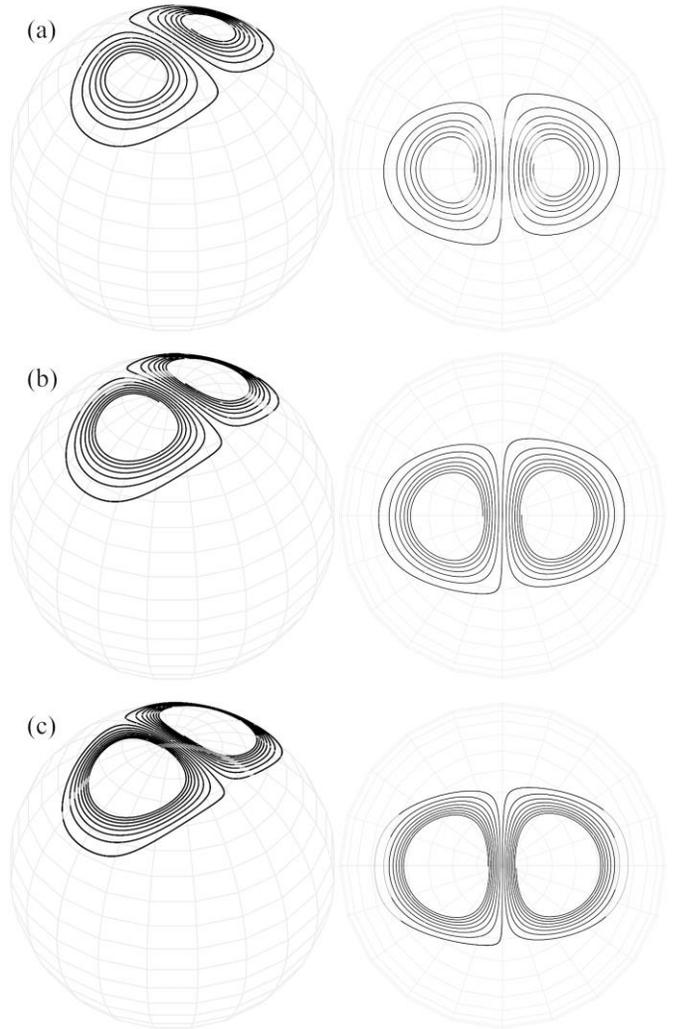

Fig. 2. Discretized conductor paths of a bent equivalent of the commercial D70 coil as an example. (a) $k = 0.08$ resulting in 12 turns and an inductivity of 5.5 μH, (b) $k = 0.04$ resulting in 14 turns and an inductivity of 9.58 μH and (c) $k = 0.02$ resulting in 16 turns and an inductivity of 15.6 μH.

We generate continuous wire representations as integral curves through a vector field which we generate from two components. The equivalent current is a divergence-free curl field. We add a small weighted divergence component to it to turn an integral curve of the superposition into a spiraling line that widely maintains the current strength and direction. The weight of the divergence component controls the number of turns and the inductance in a continuous way. The particular challenge solved here *en passant* is applicability to multi-loop coils, such as figure-of-eight coils, which need to form several center points. For matching the centers of curl and divergence components automatically and constructing the divergence component orthogonal to the current solution Hilbert space, this small additional component is generated from the equivalent current density and the normal vector of the surface as

$$\mathbf{j}_{\text{grad}}(\mathbf{r}) = \hat{\mathbf{n}}(r) \times \mathbf{j}_{\text{equiv}}(\mathbf{r}), \quad (15)$$

so that

$$\nabla \times \mathbf{j}_{\text{grad}}(\mathbf{r}) = \mathbf{0}. \quad (16)$$

Thus, the divergence field stems from the projected curl solution to represent the centers of current spirals automatically as well as accurately and allow for various spirals as, for instance, in figure-of-eight coils.

In our case, $\mathbf{j}_{\text{grad}}$ was generated as the projection onto the frame or basis on the closed surface. Normalization can optionally separate direction from strength. The integral curve leading to the overall wire-wound coil is generated through

$$\mathbf{j}_{\text{res}}(\mathbf{r}) = \mathbf{j}_{\text{equiv}}(\mathbf{r}) + k \cdot \frac{\mathbf{j}_{\text{grad}}(\mathbf{r})}{\|\mathbf{j}_{\text{grad}}(\mathbf{r})\|_2} \quad (17)$$

with $k$ controlling the spiral angle, and thus the number of turns, the resulting inductance and also the quantization granularity. The lower $k$ is, the denser the windings and the higher the discretization resolution but the higher the inductance. A major distortion through this discretization is not expected, because as Eq. (15) states, the superimposed field is curl-free and therefore unable to form any loops. Loops, however, would be required for Ampere's law and Faraday's law to generate a magnetic flux.

## VII. Examples

We validated the formalism with the example of a commercially available figure-of-eight coil (D70 Alpha Coil, Magstim, Wales UK), which may be among the most used specific implementations in the field, and a more intricate and therefore less readily available clinical coil (H4 treatment coil, Brainsway, Jerusalem, Israel).

For the D70 Alpha, we used the corresponding magnetic vector potential of the coil in the Neuroimaging Informatics Technology Initiative (NIfTI) format from Simnibs v3.2.6 as the target distribution and mapped it onto a spherical surface with a radius of 110 mm [55]. The natural harmonics of the spherical surface served as a vector basis. The region of interest inside which the fields are matched in the following example is without loss of generality a sphere with a radius of 85 mm, approximately representing the dimensions of a human brain, but only defining the region of interest, explicitly not limiting the brain to a sphere. Spherical representations with those dimensions are well established in TMS [43, 56]. In stark contrast to conventional spherical models, the outcome of this equivalency, which matches the magnetic vector potential, is not limited to this assumption, but the resulting mapped coil will be equivalent for any anatomy inside this region of interest.

The source coil resides on a planar surface so that the outer wings are relatively far from a subject's head associated with poor coupling and unnecessarily large stray flux. We demonstrate the method to map the coil onto a bent surface to reduce the necessary field energy. The scalar product (Eq. 10) with its quality of a similarity metric in combination with the underlying magnetic vector potentials inside the region of interest, specifically the sphere with a radius 85 mm here, serves to quantify the deviation introduced during numerical mapping and quantization of the example.

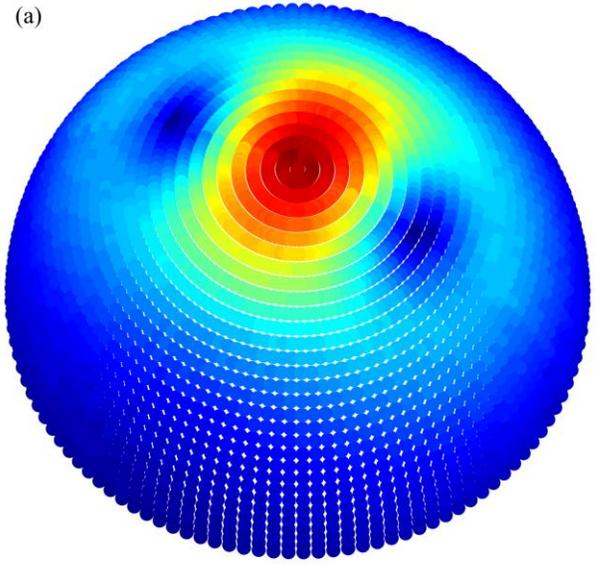

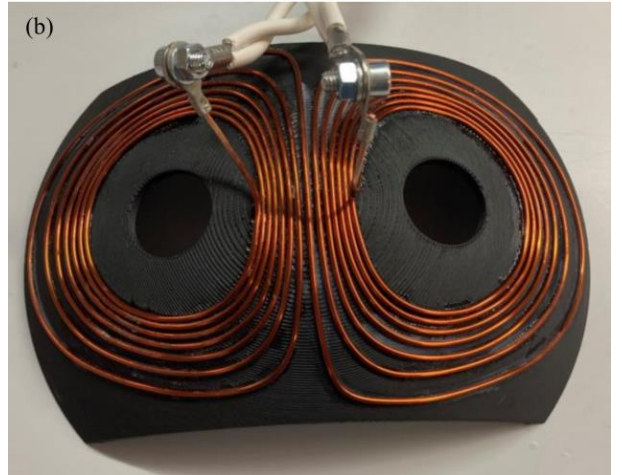

Fig. 3. Experimentally detected field and corresponding implementation of the coil from Fig. 2(b). (a) Plot of E-Field magnitude sampled under spherical constraints and (b) implemented experimental coil with 14 turns.

For comparison in one application, we introduced a realistic head model with gyrus-level anatomy into the post-hoc analysis of the model-free coil mapping. We calculated the magnetic vector potentials $\mathbf{A}$ of the three relevant conditions (calculated from the coil to be mapped, from the continuous current distribution, and from the quantized conductor path with 16 turns) and applied them with a realistic head model in Simnibs using the ernie reference model. Without loss of generality, the coils were placed on the C3 electrode in the 10–20 system with an approximately perpendicular field to the local sulcus.

To also provide a more abstract, but rather interesting example, we furthermore derived an equivalent coil based on the H4 coil. To obtain the desired magnetic vector potential from the coil, we redrew it in computer aided design (CAD) and calculated its magnetic vector potential, as well as its conductor length and inductance [72]. The boundary conditions, meaning the radius of the target surface, the radius of the sphere forming the region of interest, wherein the brain may reside, and the vector basis generated with the harmonics on the target surface for the mapping the H4 coil were the very same as for the D70 Alpha Coil.

Based on the magnetic vector potential, we calculated an equivalent current distribution discretized into a coil with ten turns to be in the usual range of inductance for TMS coils. Figure 5 displays both the original H4 coil as well as the simplified version.

## VIII. Implementation For Field Measurement

We derived equivalent coils for the D70 Alpha with various numbers of turns, providing different inductivity values (Fig. 2). Out of these, we implemented variant (b) with a factor $k$ of 0.04. We transferred the winding pattern onto a spherical polymer former and wound the coil with 3.28 m of 2 mm² magnet wire. The coil winding was connected to two 10 AWG braided cables, resulting in a cross-section of 10.5 mm² (Alpha Wire 391045, Elizabeth (NJ), USA) and an Anderson connector (Anderson Power SB-350, Ideal Industries, Sycamore (IL), USA). The coil with cable and connector was measured at 10.8 µH (Hameg Instruments HM8118, Frankfurt, Germany).

We sampled the induced electric field of the coil under spherical constraints, as seen in Fig. 3, using a well-established automated field probe that is widely used in TMS technology labs around the globe and described in more detail in the literature [43]. The coil was connected to a commercial MagVenture MagLite pulse source (Tonica, Farum, Denmark) with biphasic pulses (standard mode).

Simulations in realistic head models with gyrus-level anatomic details were performed in SimNIBS v3.2.6 with the ernie reference model.

## IX. Results For The Examples

Figure 1 displays different current distributions for increasing orders of the basis vectors. Accordingly, the detail level of the projection grows with the order of included base vectors. At this point, the coil is still a continuous vector field, which needs a discretized representation in a wire path.

Further processing of the current distribution from the highest possible resolution results in the discretized conductor paths through Equations 11 and 12. Figure 2 displays various discretizations of the simplified D70 Alpha equivalent with different factors $k$, which further entail different numbers of turns and inductances. Comparable to other discretization processes, the necessary quantization of the current density into a conductor path introduces slight deviations from the continuous shape. Accordingly, the magnetic vector potential and the primary electric field matching amount to $\gamma_{12T} = 96\%$ for the 12-turn figure-of-eight coil in Figure 2(a), $\gamma_{14T} = 98.7\%$ for the 14-turn coil in Figure 2(b), and $\gamma_{16T} = 99.4\%$ for the 16-turn coil in Figure 2(c), demonstrating an increasing similarity with the target coil's field for growing quantization resolution of the discrete wire representation. Further indicators qualifying the matching and the numerical implementation are the half-depth and tangential spread values: The D70 coil has a $d_{1/2}$ of 1.5 cm and a $s_{1/2}$ of 15 cm². The projected coil has a $d_{1/2}$ of 1.5 cm and a $s_{1/2}$ of 16 cm².

Figure 3 provides in Panel (a) the electric field measured from the experimental prototype under spherical constraints of the implementation of the coil shown in (b) matching the commercial D70 Alpha coil.

As can be seen in Figure 4, all three figure-of-eight coil representations (D70, mapped coil with continuous current distri-bution, and mapped coil with discretized turns) lead to comparable induced electric field distribution, depth penetration, and strength with negligible deviations in a realistic head model. The largest field strengths evolve at the cortical target with comparable area and roll-off. These electric field distributions were calculated independently from the derivation of the coils and have therefore no contribution to it. They are only meant to provide a demonstration of the equality of the electric fields. Despite the corresponding field distribution of both coils, the matched figure-of-eight coil on the curved spherical surface only requires 77% of the current of the original D70 coil for comparable field strength in the cortex. The coil heating, which scales with the squared current and the conductor length for equal cross section, decreases accordingly by −39%. Due to equal inductances, the required field energy of the mapped coil furthermore amounts to only 60% of the original D70, illustrating the achievable gain of such mapping.

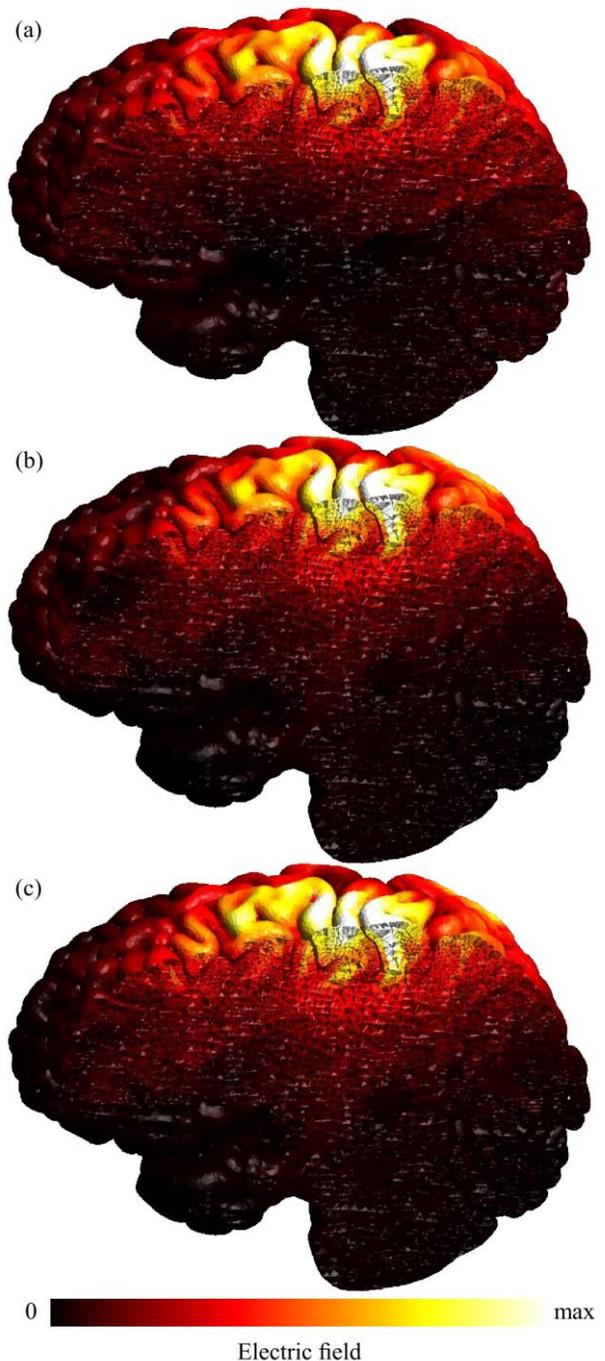

Fig. 4. Simulation of the electric field in SimNIBS using the ernie reference head model. (a) Induced electric field of the commercial Magstim D70 alpha coil to be mapped, (b) induced electric field of the matched equivalent with continuous current distribution, and (c) induced electric field of the matched coil with discretized conductors with 14 turns. The field magnitude is color-coded.

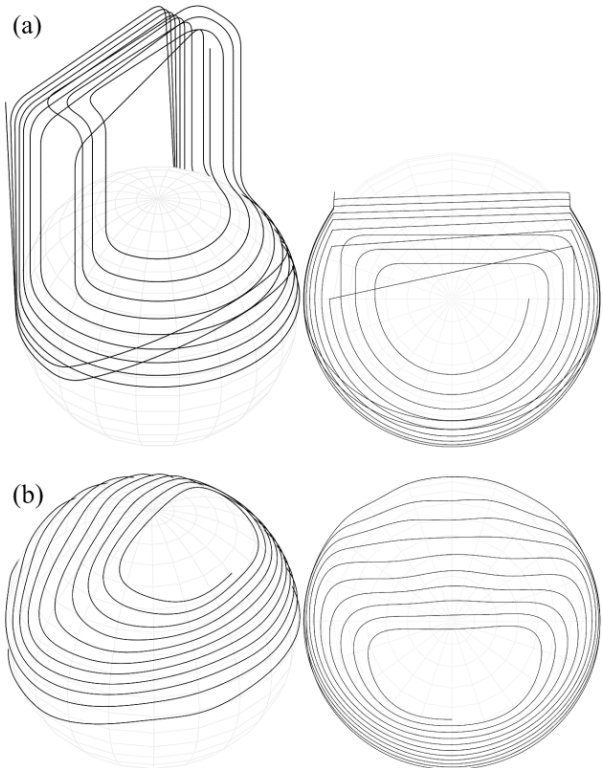

Fig. 5. Conductor paths of (a) the H4 Coil with an inductance of 31.4 μH and a conductor length of 9.1 m and (b) a spherical equivalent coil with an inductance of 13.8 μH and a wire length of 4.9 m.

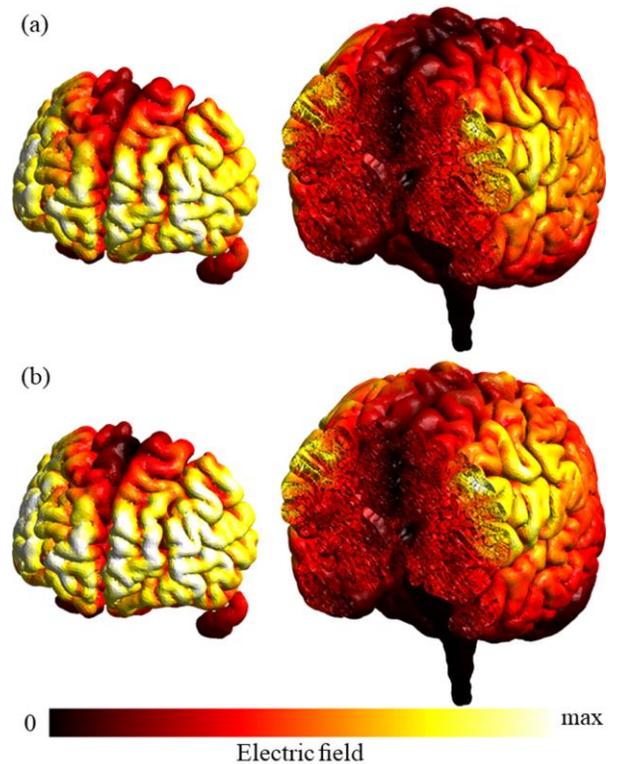

Fig. 6. Induced electric fields in a realistic anatomy in SimNIBS using the ernie reference head model with a frontal section through the insular cortex. (a) Original H4 coil to be mapped, (b) induced electric field of the simplified coil.

Figure 5 displays the original H4 and its simplified, surface-projected version. Figure 6 in turn presents the coils' electric fields, which were calculated based on their magnetic vector potentials with the ernie anatomy. The field images present the surface as well as a coronal section near the insular cortex to represent the depth properties in the main treatment area. Compared to the figure-of-eight coil (see Fig. 4), the H4 field intentionally covers a wide range of the cortex and also reaches notably deeper. Visual inspection indicates a strong similarity of the original coil and its simplified projection. The more important quantitative error between the original H4 and the simplification amounts to 4.1%. In addition to the simplification of the wiring, which intends to reduce cost and make the coil available to a wider community, the coil further also improves important practical properties: While maintaining the main characteristics of the original coil, the equivalent coil only requires 44% of the magnetic field energy for the same induced electric field and 54% of the coil wire, cutting the heating losses by 46% at comparable stimulation amplitude.

## X. Discussion

The mapping of the reference coils' **A** vector potentials onto a bent surface reduced the necessary field energy and reflects previous observations that efficient coils should be as close to the head as possible while avoiding elements standing off [5, 57-59, 73]. The lower magnetic field energy levels and currents for the projected coils would not only allow shrinking the TMS device size but also promises to reduce the coil heating and coil sound, both proportional to the squared current, as two technical side effects limiting the application of TMS [60-64].

The matching of both the D70 figure-of-eight coil as well as of the clinical H4 coil in the region of interest was close to one. The remaining deviation results from the quantization of the coils into a finite number of turns and are on the order of the manufacturing tolerance of some coils [43].

The proposed procedure fills a number of blank spots in the prior art. Previous matching procedures left gaps in the process, which had to be performed manually with often rather ad-hoc steps: First, the matching was ensured for one or few specific anatomies only as the highly individual induced electric field was in the focus, which depends on the material properties and features such as the gyrification [65, 66, 67, 73]. We instead base the procedure on an anatomy-independent precursor to achieve general matching. For anatomy-specific coil optimization or matching, this step can certainly also follow the previous approaches and use the induced electric field together with the other steps.

Furthermore, the presented method introduces a consistent Hilbert-space-based formalism for a projection or basis change instead of previous computationally expensive iterative optimization, which furthermore tends to turn field or coil matching into an open-ended iterative search [73]. The target surface does not need to be between the source coil and the region of interest but can also be larger. However, as distance for a coil has low-pass-filtering properties, the closer the target surface is to the head, the better it can represent fine features, such as tight bends. The coil space and the space of the magnetic vector potential were designed as Hilbert vector spaces and equipped with vector bases as well as an isomorphism between the spaces. Whereas previous optimization-based matching could not guarantee to find a good match for design reasons but only hope for both good convergence and a sufficiently general solution space, the basis change maps a coil basis vector by basis vector so that it can guarantee a match and also quantify (through the inner product) general, anatomy-independent matching.

Finally, the prior art did not offer a satisfactory solution for generating a quantized implementable coil with individual turns in a spiraling fashion. Previously, any discretization of continuous vector fields into conductor paths from precursors of this work followed two steps: first, closed wire loops were discretized into isolines of the field; second, the individual closed loops representing individual turns were opened in one place and manually either spatially smoothed into each other or connected in suitable spots [68]. This technique results in rather abrupt steps between the turns of the conductor path and requires many manual steps, especially for complex geometries.

As a solution, we introduced a coherent method that intrinsically generates the spiraling with a line integral through a vector field that combines basis superposition with a corresponding divergence field formed out of the latter. This method can intrinsically manage multiple loop centers. It was therefore intentionally tested on a figure-of-eight coil as it contains two centers and therefore requires a very general and robust method. Furthermore, the solution sets the number of turns and thus the inductance and the discretization resolution naturally with a single parameter that defines the strength of the divergence field. This single parameter can continuously and monotonically control the inductance so that the method can easily match a target inductance. In oscillator circuits of conventional TMS devices without active pulse control, the target inductance controls the pulse duration. Previous manual search of an inductance with connecting contour lines can be a rather tedious iterative procedure.

The method was designed for general bases and frames. The specific example used harmonics on the coil surface as in previous work. However, harmonics are nonlocal, whereas coils are typically spatially well-defined objects so that higher orders are not only necessary for describing sharper features of a coil but also to shift the coil and compensate for errors of lower orders. Bases with more local dominance or even compact carriers might deserve more research effort and may show better numerical stability and convergence in the future.

## XI. Conclusion

In contrast to previous optimization approaches to artificially generate entirely novel coils and fields, this article presents a new formalism to generate equivalents to exploit the potential of existing but often limited coils, regularly approved for procedures through their field shape, on given different geometric surface using Huygens' and Love's principle. These well-known principle state that an electric field distribution inside a volume can be generated by a current distribution on an arbitrarily shaped shell around this volume. The method is based on a modal decomposition of the current density as well as the corresponding magnetic vector potential. After achieving an equivalent current distribution on the surface, the formalism continues by modifying this distribution by an overlaid gradient field and discretizing it with a line integral into an actually usable conductor path.

These benefits not only make it easy to use because no manual steps are required, but it is comparatively also less computationally intensive, because of the missing optimization process and can therefore be easily performed on midrange consumer hardware.

Among others, the presented formalism allows mapping coils based on their magnetic vector potentials onto surfaces that are closer to the head to save energy, e.g., to only one third in the presented D70 example and 44% for the H4 coil. Particularly, it allows mapping flat, or even complicated 3D-shaped coils onto bent surfaces or even the head surface in a fast forward formalism that avoids any previous computationally limited iterative fitting procedures. It furthermore can derive specific coil implementations for coils for which only external field measurements are available. In addition, the method can also *re-trace* coils with different numbers of turns or other geometric parameters, such as the surface shape on which the coil resides, the inner diameter, outer diameter, or coil height. Similar to measured field distribution, the method can further generate the closest coil that generates a user-defined field distribution independent of already existing coil geometries.